\documentclass[fleqn,usenatbib]{mnras}

\usepackage[T1]{fontenc}

\DeclareRobustCommand{\VAN}[3]{#2}
\let\VANthebibliography\thebibliography
\def\thebibliography{\DeclareRobustCommand{\VAN}[3]{##3}\VANthebibliography}

\usepackage[dvipdfmx]{graphicx}	
\usepackage{amsmath}	
\usepackage{amssymb}	
\usepackage[utf8x]{inputenc}
\usepackage[dvipdfmx]{color}
\usepackage{color}
\usepackage{bm}
\usepackage{color}
\usepackage{colortbl}
\usepackage{tikz}
\usetikzlibrary{trees}

\title[Scale decomposition of angular momentum transport]
{Investigation of the dependence of angular momentum transport on spatial scales for construction of differential rotation}

\author[K. Mori and H. Hotta]{
K. Mori$^{1}$\thanks{E-mail: mn@ras.org.uk (KTS)} and
H. Hotta$^{1}$
\\
$^{1}$Department of Physics, Graduate School of Science, Chiba University 1-33 Yayoi-cho, Inage-ku, Chiba 263-8522 Japan\\
}

\date{Accepted XXX. Received YYY; in original form ZZZ}

\pubyear{2022}

\begin{document}
\label{firstpage}
\pagerange{\pageref{firstpage}--\pageref{lastpage}}
\maketitle

\begin{abstract}
  We investigate the dependence of the angular momentum transport (AMT) on the spatial scales with numerical simulation of solar-like stars. It is thought that turbulence has an essential role in constructing solar differential rotation (DR). In a widely used method to analyse the construction mechanism of DR, the flow is divided into two components, `mean flow' and `turbulence', where `turbulence' includes a broad spectrum of spatial scales. The features of the AMT are expected to depend on the scale. In this study, we decompose the angular momentum flux (AMF) to investigate the dependence of the AMF on the spatial scale. We compare the results with anti-solar- (fast pole) and solar-type (fast equator) DR. Our conclusions are summarized as 1. Radially outward AMT is seen on a large scale ($60~\mathrm{Mm}\le L<120~\mathrm{Mm}$) in rotationally constrained systems. 2. Even when the scale-integrated AMF is negative, we sometimes observe positive AMF on certain scales. 3. Small-scale turbulence tends to transport the angular momentum radially inward and causes the anti-solar DR, indicating that high-resolution simulation is a negative factor for solar-like DR. Our method to decompose the AMF provides a deep understanding of the angular momentum and construction mechanism of DR.
\end{abstract}




\section{Introduction}
\label{sec:introduction}
We can observe differential rotation (DR) in the solar convection zone, in which the rotation rate depends on the latitude.
Helioseismology has revealed the detailed profile of the DR \citep{schou_1998ApJ...505..390S}. In the solar case, the equator rotates faster and the polar region rotates slower; this DR is called solar-type DR. Conversely, DR with a slow equator and fast poles is called anti-solar-type DR.
\par
Turbulent thermal convection is thought to form the DR. The Coriolis force induced by the solar rotation makes the solar turbulent thermal convection anisotropic. Consequently, turbulence transports the angular momentum in a specific direction, which leads to large-scale flow construction.
This transport generates large-scale flow, i.e. DR and meridional flow (MF).
\par
We briefly summarize the mechanism of constructing large-scale flow by turbulent angular momentum transport (AMT). We start with the angular momentum conservation equation derived from the longitudinal equation of motion in an inertial (non-rotating) frame in the spherical geometry ($r,\theta,\phi$):
\begin{align}
	\frac{\partial}{\partial t}(\rho_0 \langle\mathcal{L}\rangle) = -\nabla\cdot(\rho_0 \lambda \langle u_\phi\bm{u}_\mathrm{m}\rangle),
\end{align}
where $\rho$ is the density. $\bm{u}_\mathrm{m}=u_r\bm{e}_r + u_\theta\bm{e}_\theta$ and $u_\phi$ are the meridional and longitudinal fluid velocities in an inertial system\footnote{We use $\bm{v}$ for the fluid velocity in a rotating frame}.
$\bm{e}_r$ and $\bm{e}_\theta$ are the unit vectors in the radial and the latitudinal directions, respectively. $\lambda=r\sin\theta$ is the distance from the rotational axis. $\mathcal{L}=\lambda u_\phi$ is the specific angular momentum. The subscript 0 for the density indicates the spherically symmetric value. $\langle\rangle$ means the longitudinal average. We note that we ignore the magnetic field and viscosity in this discussion for simplicity.
Typically, we divide the fluid velocity $\bm{u}$ into the mean part $\langle\bm{u}\rangle$ and the turbulent (perturbed) part $\bm{u}'$, i.e. $\bm{u}=\langle \bm{u}\rangle + \bm{u}'$. Then, we obtain the angular momentum conservation equation with mean flow and turbulence contributions:
\begin{align}
  \frac{\partial}{\partial t}(\rho_0\langle \mathcal{L} \rangle) = 
  &\underbrace{-\nabla\cdot\left(\rho_0\langle \mathcal{L} \rangle\langle\bm{v}_{\mathrm{m}}\rangle\right)}_{\mbox{mean flow contribution}} \nonumber\\
  &\underbrace{
  - \nabla\cdot\left(\rho_0\lambda\langle\bm{v}'_m v'_{\phi}\rangle\right).
  }_{\mbox{turbulence contribution}}
	\label{angmoment}
\end{align}
When the turbulence is anisotropic, the term $\langle \bm{v}_{\mathrm{m}}' v_{\phi}'\rangle$ on the right side becomes important and transports angular momentum in specific directions. We note that $\bm{u}_\mathrm{m}=\bm{v}_m$ and $u'_\phi=v'_\phi$.
This procedure is broadly used to understand the construction mechanism of DR \cite[e.g.][]{brun_2002ApJ...570..865B,hotta_2015ApJ...798...51H}.
\par
One may notice that the division of `mean flow $\langle\bm{v}\rangle$' and `turbulence $\bm{v}'$' is a somewhat rough idea. `Mean flow' is the longitudinal average in this discussion. This definition is reasonable because we are usually interested in the DR and MF observations, which are defined with the longitudinal average. `Turbulence' in the discussion has a broad meaning because it includes a broad spectrum of spatial scales except for the longitudinal average. There should be a more detailed scale dependence of the AMT by turbulence.
\par
It is known that the influence of the rotation changes the direction of the radial AMT. Roughly speaking, a weak (strong) influence of the rotation leads to a negative (positive) radial AMT \citep[e.g.][]{karak_2015A&A...576A..26K}. \cite{featherstone_2015ApJ...804...67F} suggest that a negative radial AMF and resulting one-cell MF are the essential reasons for the anti-solar DR. In addition, it is known that higher-resolution simulations more easily fall into the anti-solar DR than lower-resolution ones \citep[e.g.][]{hotta_2015ApJ...798...51H}. This result indicates that the small-scale turbulence introduced in the high-resolution simulations transports the angular momentum radially inward. To confirm this tendency, we need to investigate the scale dependence of the AMF, i.e. the scale decomposition of the AMF is required.
\par
In this study, we decompose the AMT by the Reynolds stress (turbulence), i.e. $\langle \bm{v}'_\mathrm{m} v'_\phi\rangle$, on different scales. We analyse the results of three-dimensional magnetohydrodynamic numerical simulations in a spherical shell.
We investigate the scale dependence of the AMT, and the key questions are:
\begin{enumerate}
	\item How does the AMT depend on the spatial scale?
	\item Does small-scale turbulence tend to transport the angular momentum radially inward?
\end{enumerate}
This manuscript is constructed as follows:
We describe the model setup for the numerical simulations in section \ref{sec:model}. Our proposed method to decompose the AMT is introduced in section \ref{sec:analysis}. We present the simulation results and explain the process of DR formation in section \ref{sec:result}. The scale decomposition of the AMF is also shown in section \ref{sec:result}. Finally, we summarize and conclude the paper in section \ref{sec:conclusion}.
\section{Numerical Model}
\label{sec:model}
We explain our numerical model setup in this section. We solve the three-dimensional magnetohydrodynamic equations in the spherical geometry $(r, \theta, \phi)$. The whole sphere is covered with the Yin--Yang grid \citep{kageyama_2004GGG.....5.9005K}.
The radial extent of the computational domain is $0.71R_{\odot} \leq r \leq 0.96R_{\odot}$, where $R_\odot$ is the solar radius. The number of grid points is $(N_r, N_{\odot}, N_{\phi}, N_{\mathrm{YY}}) = (256 \times 256 \times 768 \times 2)$, where $N_r$, $N_\theta$, and $N_\phi$ indicate the grid points in radial, latitudinal, and longitudinal directions, respectively. $N_\mathrm{YY}$ is the factor for the Yin--Yang grid. For our analyses, we convert the Yin--Yang grid to ordinal spherical geometry with the number of grid points of $(N_r, N_{\theta}, N_{\phi}) = (256 \times 512 \times 1024)$.
We use the R2D2 (Radiation and RSST for Deep Dynamics) code \citep{hotta_2019SciA....5.2307H,hotta_2020MNRAS.494.2523H,hotta_2021NatAs...5.1100H}. The magnetohydrodynamic equations are expressed as:
\begin{align}
  &\frac{\partial\rho_1}{\partial t} = -\frac{1}{\xi^2}\nabla\cdot(\rho\bm{v}),
  \label{mass_flux} \\
  &\frac{\partial}{\partial t}(\rho\bm{v}) = -\nabla\cdot(\rho\bm{vv}) - \nabla p_1 - \rho_1g\bm{e}_r \nonumber \\
  &\hspace*{1.5cm} + 2\rho\bm{v}\times\bm{\Omega}_0 + \frac{1}{4\pi}(\nabla\times\bm{B})\times\bm{B},
  \label{eq_of_motion}\\
  &\frac{\partial\bm{B}}{\partial t} = \nabla\times(\bm{v}\times\bm{B}), \\
  &\rho T\frac{\partial s_1}{\partial t} = -\rho T(\bm{v}\cdot\nabla)s + Q_s,
  \label{entropy} \\
  &p_1 = \left(\frac{\partial p}{\partial \rho}\right)_s\rho_1 + \left(\frac{\partial p}{\partial s}\right)_{\rho}s_1,
  \label{p_1}
\end{align}
where $\rho, \bm{v}, \bm{B}, s,$ and $p$ are density, velocity, magnetic field, specific entropy, and pressure, respectively. The subscripts 0 and 1 express the spherical symmetric background and the perturbation, respectively. $\xi$ is the factor for the reduced speed of sound technique \cite[RSST][]{hotta_2012A&A...539A..30H} to relax the Courant--Friedrich--Lewy (CFL) condition by the fast sound wave. The effective sound speed is reduced by a factor $\xi$, and we fix the reduced speed of sound at $2.5~\mathrm{km~s^{-1}}$.
Because we deal with small perturbations of $\rho_1/\rho_0 \sim p_1/p_0 \sim T_1/T_0 \sim 10^{-6}$, we use the linearized equation of state (eq.~(\ref{p_1})). The coefficients $(\partial p/\partial\rho)_s$ and $(\partial p/\partial s)_{\rho}$ are calculated with the OPAL repository \cite[]{rogers_1996ApJ...456..902R}.
We use the Model S \citep{christensen_1996Sci...272.1286C} for the background stratification and related variables \cite[see][for more details]{hotta_2020MNRAS.494.2523H}.
$\bm{\Omega_0}$ is the angular velocity vector of the system, where $\bm{\Omega_0} = \Omega_0(\cos\theta\bm{e}_r - \sin\theta\bm{e}_{\theta})$. We prepare three cases, $\Omega_0 = 1\Omega_{\odot}$, $2\Omega_{\odot}$, and $3\Omega_{\odot}$, where $\Omega_{\odot}/2\pi = 413\ \mathrm{nHz}$, $\Omega_{\odot}$ is the solar reference angular velocity.
We use the heating term $Q_s$ defined as:
\begin{eqnarray}
  &&Q_s = \frac{1}{r^2}\frac{\partial}{\partial r}[r^2(F_{\mathrm{rad}} + F_{\mathrm{art}})], \\
  &&F_{\mathrm{rad}} = -\kappa_r\frac{dT_0}{dr}, \\
  &&F_{\mathrm{art}} = \frac{L_{\odot}}{4\pi r^2_\mathrm{max}}\left(\frac{r}{r_{\mathrm{max}}}\right)^2\mathrm{exp}\left[-\left(\frac{r - r_{\mathrm{max}}}{d_{\mathrm{art}}}\right) \right],
\end{eqnarray}
where $F_{\mathrm{rad}}$ and $F_\mathrm{art}$ are the radiation flux and artificial energy flux. Because our calculation does not include the photosphere, we need artificial cooling $F_\mathrm{art}$ to drive the thermal convection. We extract the solar luminosity $L_{\odot}$ at the upper boundary $r_{\mathrm{max}} = 0.96R_{\odot}$. We use the diffusion approximation for the radiation energy transfer $F_\mathrm{rad}$. The radiative diffusion coefficient $\kappa_\mathrm{r}$ is also obtained from the OPAL repository.
$d_{\mathrm{art}}$ is the depth of the artificial cooling layer of $d_{\mathrm{art}} = 2H_p(r_{\mathrm{max}})$, where $H_p(r_\mathrm{max}) = 9.46~\mathrm{Mm}$ is the pressure scale height at $r = 0.96R_{\odot}$.
We continue the calculation for $5000$ days for each case, and the following results are averaged between $t=4500$ to $5000$ days unless otherwise noted.
\section{Method of decomposition}
\label{sec:analysis}
In this study, we decompose the turbulence angular momentum flux (AMF) to scale-dependent values. Following the discussion around eq. (\ref{angmoment}), the AMF by the Reynolds stress can be written as
\begin{align}
  F_{\mathrm{R},\alpha} = \rho_0\lambda\langle u'_\alpha u'_\phi\rangle,
\end{align}
where $\alpha=r$ or $\theta$.
We decompose $F_{\mathrm{R}, \alpha}$ with the Fourier transform in the longitudinal direction and Parseval's theorem.\par
We discuss a physical quantity $g$ defined at each longitudinal position $\phi$. We define the Fourier transform in the longitudinal direction as:
\begin{align}
  \widehat{g}(m) = \frac{1}{2\pi}\int_0^{2\pi} g(\phi)\exp\left(-im\phi\right)d\phi,
  \label{fourier}
\end{align}
where $m$ is azimuthal order or the non-dimensional wave number in the longitudinal direction. The wavelength of each mode is $L_m=2\pi\lambda/m$. Parseval's theorem implies that the correlation of two quantities ($g_1(\phi)$ and $g_2(\phi)$) can be expanded with two separately Fourier-transformed values ($\widehat{g}_1(m)$ and $\widehat{g}_2(m)$) as
\begin{align}
  \langle g_1(\phi)g_2(\phi)\rangle = \sum_{m=0}^{N-1} \widehat{g}_1(m)\widehat{g}_2^*(m),
  \label{pars_org}
\end{align}
where $^*$ indicates the complex conjugate. This relation shows that $\widehat{g}_1(m)\widehat{g}_2^*(m)$ is the contribution of wave number $m$ to the correlation $\langle g_1(\phi)g_2(\phi)\rangle$. Because the numerical result is discretized data, we relate the Fourier transform shown in eq. (\ref{fourier}) to the discrete Fourier transform (DFT) as:
\begin{align}
  \widehat{g}(m) \sim \frac{1}{N_\phi}\sum_{k = 0}^{N_\phi - 1}g_k\mathrm{exp}\left(-i\frac{2\pi mk}{N_\phi}\right),
  \label{dft}
\end{align}
where $g_k = g(\phi_k)$ and $\phi_k=k\Delta \phi$. $\Delta \phi = 2\pi/N_\phi$ is the grid spacing in the longitudinal direction.
Because the Reynolds stress only includes the perturbation from the longitudinal average ($m=0$), we only deal with the $m\neq0$ mode as
\begin{align}
  \langle g_1' g_2' \rangle = 
  2\sum_{m = 1}^{N_\phi/2 - 1}\mathrm{Re}[\widehat{g}_1(m)\widehat{g}^*_2(m)] + \mathrm{Re}\left[\widehat{g}_1\left(\frac{N_{\phi}}{2}\right)\widehat{g}^*_2\left(\frac{N_{\phi}}{2}\right)\right],
  \label{pars_2}
\end{align}
where $\mathrm{Re}[~]$ represents the real part of each quantity, and $g_1', g_2'$ represent perturbation from longitudinal average. Eq. (\ref{pars_2}) can be derived from eq. (\ref{pars_org}) using $g = \langle g\rangle + g' = \widehat{g}(0) + g'$. We have a factor of 2 in front of the sum because the sum is only over positive values of $m$.

\par
Because the wavelength is expressed as $L_m=2\pi\lambda/m$ with $\lambda=r\sin\theta$, the same $m$ does not always mean the same spatial scale in different positions $(r,\theta)$. Thus, we divide the Reynolds stress into four components based on the actual scale $L_m$, not on the wave number $m$ itself. Then, we collect a similar wavelength to an AMF as:
\begin{align}
  F_{\mathrm{R},\alpha} = \rho_0\lambda \langle v'_\alpha v'_\phi\rangle = \sum_{i=1}^4 f^i_{R,\alpha},~~(\alpha=r,\theta),
\end{align}
where the scale-dependent momentum flux $f^i_{R,\alpha}$ is defined as
\begin{align}
  f^i_{R,\alpha}(r,\theta) = 
  2\rho_0 \lambda\sum_{m=m_{i\mathrm{(min)}}}^{m_{i\mathrm{(max)}}} \mathrm{Re}
  \left[\widehat{v}_\alpha \widehat{v}^*_\phi\right].
\end{align}
When $m_{i(\mathrm{max})}=N_\phi/2$, the expression becomes
\begin{align}
  f^i_{R,\alpha}(r,\theta) = \rho_0 \lambda
  \left(
  2\sum_{m=m_{i\mathrm{(min)}}}^{m_{i\mathrm{(max)}}-1}\mathrm{Re}
  \left[\widehat{v}_\alpha \widehat{v}^*_\phi\right]  \right. \nonumber\\
  \left.+\mathrm{Re}\left[\widehat{v}_\alpha\left(\frac{N_\phi}{2}\right)\widehat{v}_\phi\left(\frac{N_\phi}{2}\right)\right]
  \right).
\end{align}
$m_{i\mathrm{(min)}}$ and $m_{i\mathrm{(max)}}$ are calculated with the maximum $L_{i\mathrm{(max)}}$ and minimum $L_{i\mathrm{(min)}}$ scale length included in the collection of the AMF, respectively, as
\begin{align}
  m_{i\mathrm{(min)}}(r,\theta) &= \mathrm{floor}\left(\frac{2\pi\lambda}{L_{i\mathrm{(max)}}}\right) + 1, \\
  m_{i\mathrm{(max)}}(r,\theta) &= \mathrm{floor}\left(\frac{2\pi\lambda}{L_{i\mathrm{(min)}}}\right), 
\end{align}
where $\mathrm{floor}()$ is the floor function. Our choice of $L_{i\mathrm{(max)}}$ and $L_{i\mathrm{(min)}}$ is shown in Table \ref{ta:l}.
\begin{table}
  \centering
  \caption{Our choice of $L_{i\mathrm{(max)}}$ and $L_{i\mathrm{(min)}}$ to determine $m_{i\mathrm{(min)}}$ and $m_{i\mathrm{(max)}}$ and resulting $f^i_{\mathrm{R},\alpha}$is shown. $R_{\odot}$ is the solar radius.}
  \label{ta:l}
  \begin{tabular}{lcc}
    \hline 
    $i$ & $L_{i\mathrm{(min)}}$ & $L_{i\mathrm{(max)}}$ \\
    \hline
    1 & 240~Mm & $2\pi R_{\odot}$ \\
    2 & 120~Mm & 240~Mm\\
    3 & 60~Mm  & 120~Mm\\
    4 & $4\pi\lambda/N_\phi $ & 60~Mm \\
    \hline
  \end{tabular}
\end{table}
Our definition of the scale-dependent AMF $f^i_{\mathrm{R},\alpha}$ is the AMF contribution from the scales $L_{i\mathrm{(min)}}\leq L_m< L_{i\mathrm{(max)}}$. For $i=4$, the scale-dependent AMF includes all the contributions below $60~\mathrm{Mm}$.
If $m_{i(\mathrm{min})} > m_{i(\mathrm{max})}$ at a particular position, we set $f^i_{R,\alpha} = 0$ at that position.
\section{RESULT}
\label{sec:result}
\subsection{General properties}
Table \ref{ta:ro} shows the Rossby number Ro evaluated in cases.
Ro is a dimensionless number that represents the effect of rotation and is defined as
\begin{eqnarray}
  \mathrm{Ro} = \frac{v_{\mathrm{RMS}}}{2\Omega_0 d},
\end{eqnarray}
where $d$ is the radial extent of the computational domain, and $v_{\mathrm{RMS}}$ is the rms velocity averaged in the longitudional direction.
We follow \cite{featherstone_2015ApJ...804...67F} for the definition of Ro.
\begin{table}
  \centering
  \caption{Rossby number Ro for each case.}
  \label{ta:ro}
  \begin{tabular}{lccc}
    \hline 
    case & $1\Omega_{\odot}$ & $2\Omega_{\odot}$ & $3\Omega_{\odot}$ \\
    \hline
    Ro & 0.115 & 0.059 & 0.035 \\
    \hline
  \end{tabular}
\end{table}

\begin{figure*}
	\begin{center}			
		\includegraphics[clip, width = 0.95\textwidth]{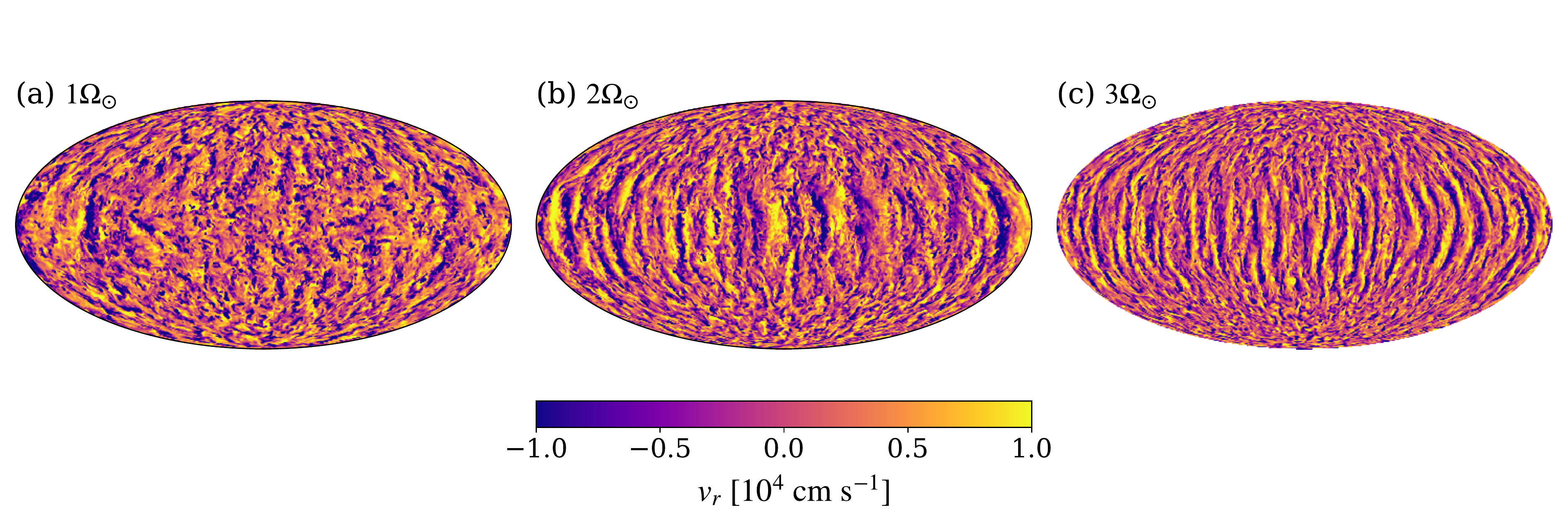}
    \caption{Radial velocity $(v_r)$ at $r = 0.83R_{\odot}$ is shown. Panels a, b, and c show the results from $\Omega_0=1\Omega_{\odot}$, $2\Omega_{\odot}$, and $3\Omega_{\odot}$ cases, respectively.
		In the case with $\Omega_0=2\Omega_{\odot}$ and $3\Omega_{\odot}$, banana cells, i.e. north--south aligned convection cells, are clearly seen, whereas typical thermal convection structures dominate in the $1\Omega_{\odot}$ case. The values in this figure are at $t=5000$ days.}
		\label{vx_at_83}
	\end{center}
\end{figure*}

The overall structure of the radial velocity $v_r$ in the calculations at $r = 0.83R_{\odot}$ is shown in Fig.~\ref{vx_at_83}.
It is known that when the rotational influence is strong, we tend to see a banana cell, i.e. a north--south aligned convection pattern \cite[e.g.][]{miesch_2000ApJ...532..593M}, which has an important role for the AMT \cite[e.g.][]{miesch_2005LRSP....2....1M}.
We clearly observe a banana cell in the $3\Omega_\odot$ case (Fig. \ref{vx_at_83}c) and the pattern becomes indistinct in the $1\Omega_\odot$ case (Fig. \ref{vx_at_83}a). We note that the difference in the banana-cell appearance between cases is not seen in the upper layers ($r = 0.96R_{\odot}$, the figure is not shown).
\par
\begin{figure}
	\begin{center}			
		\includegraphics[clip, width=7cm]{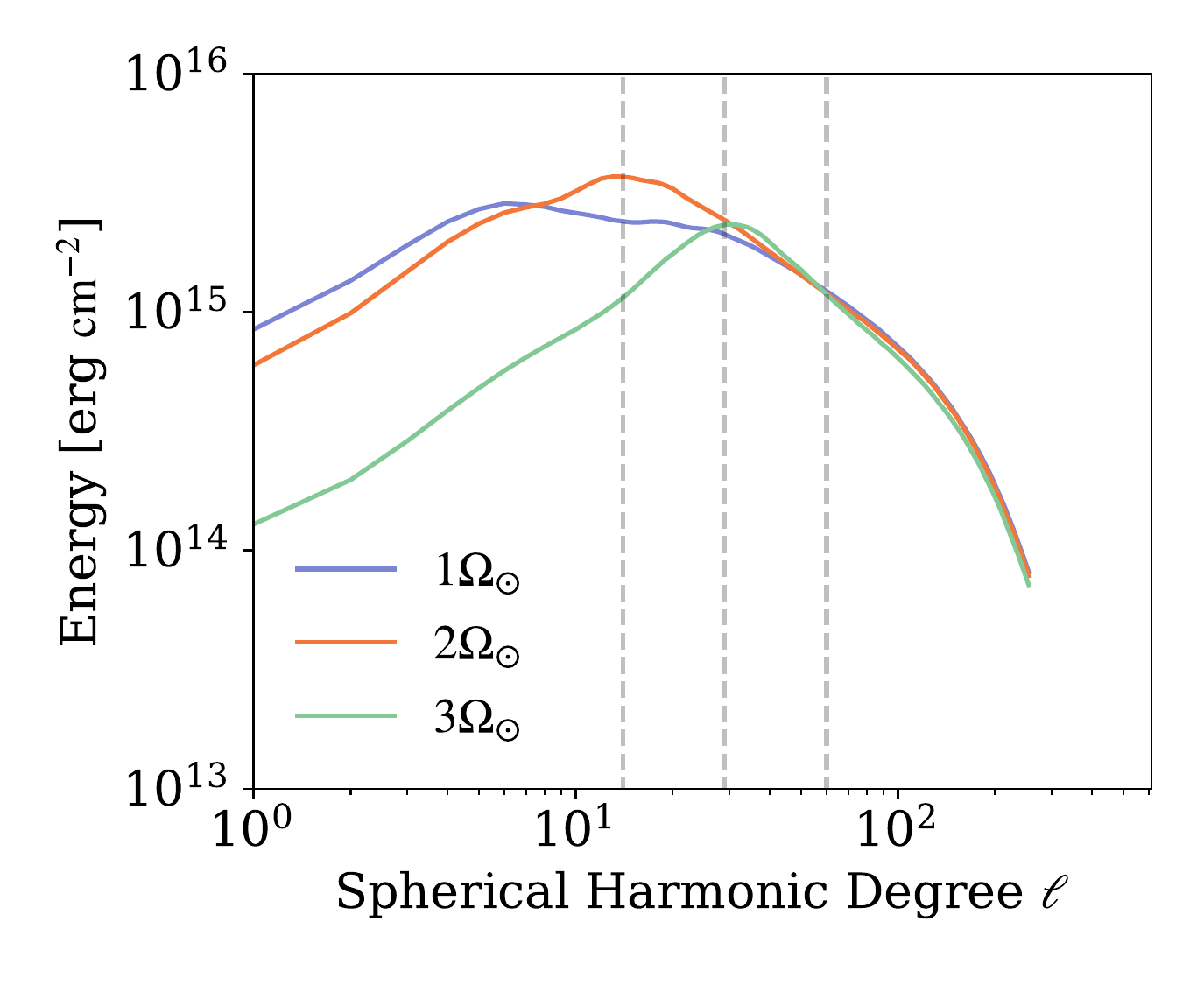}
    \caption{Kinetic energy at $r=0.83R_{\odot}$ is shown. Blue, orange, and green lines show the results from $1\Omega_\odot$, $2\Omega_\odot$, and $3\Omega_\odot$ cases, respectively. The large-scale energy ($\ell<30$) is significantly reduced in the $3\Omega_\odot$ case.
The three vertical dashed lines indicate $\ell$ corresponding to $240$, $120$, and $60\ \mathrm{Mm}$, which divide the scale-dependent AMF $f^i_{\mathrm{R},\alpha}$. A Gaussian filter with three grid points width is applied to reduce the realization noise.}
		\label{spectrum}
	\end{center}
\end{figure}

Fig. \ref{spectrum} shows the spectra of the kinetic energy. Blue, orange, and green lines show the results from $1\Omega_\odot$, $2\Omega_\odot$, and $3\Omega_\odot$ cases, respectively. We follow the definition of \cite{hotta_2022ApJ...933..199H} for the kinetic energy spectra (See their eq. (16)).
The relation between the spatial scale $L$ and the spherical harmonic degree $\ell$ is
\begin{align}
	L_\ell(r) = \frac{2\pi r}{\sqrt{\ell(\ell+1)}},
\end{align}
The large-scale kinetic energy ($\ell<30$) is suppressed in the $3\Omega_\odot$ case (green line), whereas the energy does not depend on the rotation rate $\Omega_0$ in the small scale ($\ell>30$). This indicates that the influence of rotation on the flow is reduced on the small scale.
\par
\begin{figure}
	\begin{center}			
		\includegraphics[width = 0.5\textwidth]{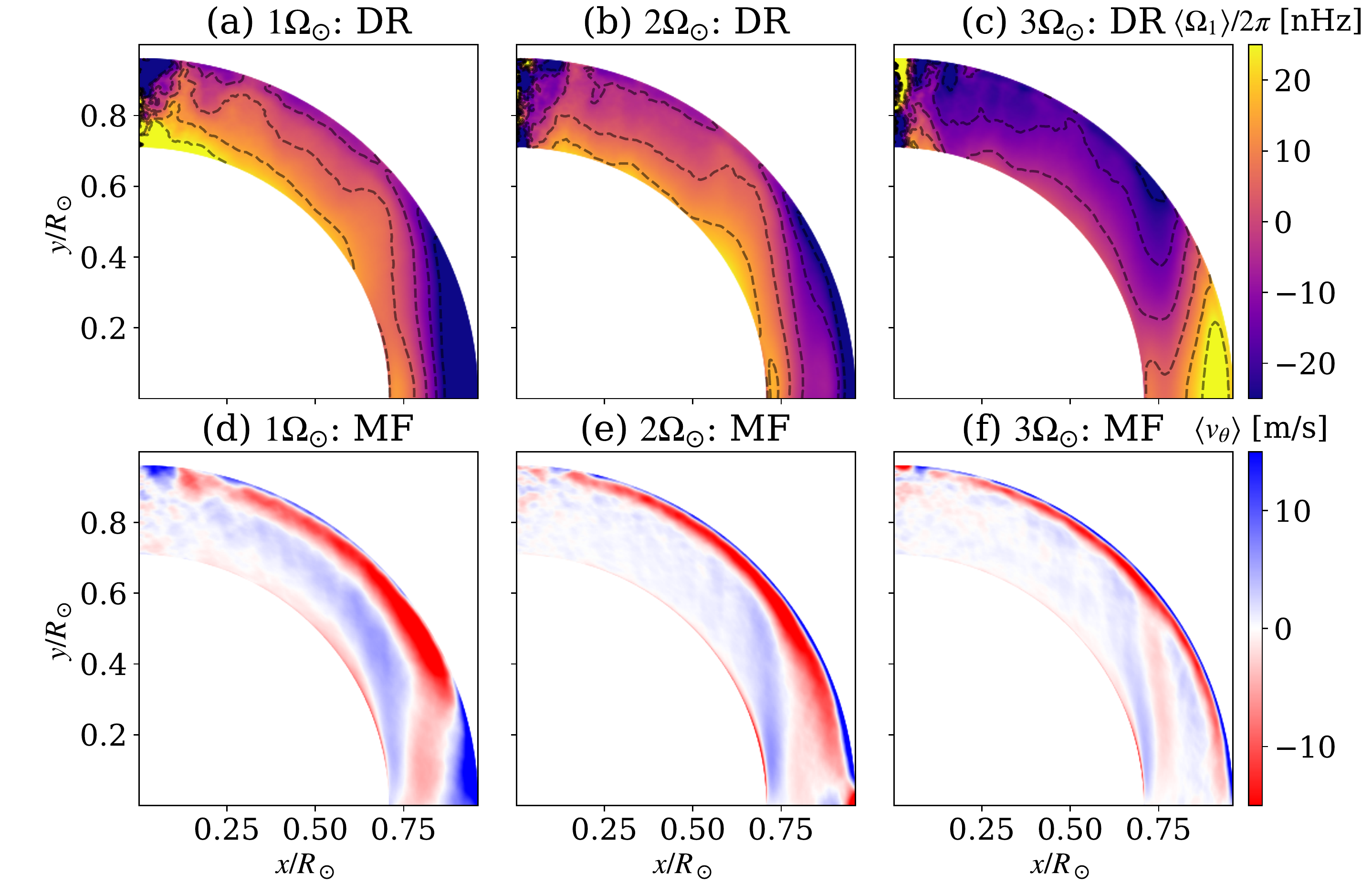}
    \caption{The DR $\langle\Omega_1\rangle/2\pi$ (upper panels), and the MF $\langle v_{\theta} \rangle$ (bottom panels), for each case are shown. The results from $1\Omega_\odot$ (panels a and d), $2\Omega_\odot$ (panels b and e), and $3\Omega_\odot$ (panels c and f) are shown.
    We can observe the anti-solar DR in the $1\Omega_{\odot}$ and $2\Omega_{\odot}$ cases, where the polar region is rotating faster, whereas the $3\Omega_\odot$ case shows the solar-like DR. As for the MF, we can see a mostly single-cell flow in the $1\Omega_{\odot}$ and $2\Omega_{\odot}$ cases, whereas a multi-cell MF is formed in the $3\Omega_{\odot}$ case.}
	\label{omgea_and_vym_all}
	\end{center}
\end{figure}

Fig. \ref{omgea_and_vym_all} shows the DR (upper panels) and the MF (bottom panels).
The $1\Omega_{\odot}$ and $2\Omega_{\odot}$ cases have anti-solar DR (panels a and b), whereas the $3\Omega_{\odot}$ case has solar-type DR (panel c). We obtain the fast pole in the $2\Omega_{\odot}$ case since we only include numerical diffusivity in our simulations. In most of simulations, explicit diffusivities, which lead to effectively lower resolution. In our simulation, small-scale turbulence is introduced due to effective high resolution. As a result, inward radial transports dominate and the polar acceleration occurs in the $2\Omega_{\odot}$ case as well as in the $1\Omega_{\odot}$ case (see also \S \ref{sec:result_2}).
Regarding the MF, $1\Omega_{\odot}$ and $2\Omega_{\odot}$ cases have a single-cell, whereas $3\Omega_{\odot}$ case has multi-cell MF in a hemisphere. This result is consistent with \cite{featherstone_2015ApJ...804...67F}. They argued that the single-cell MF transports the angular momentum poleward and tends to construct the anti-solar DR.
\par
\begin{figure}
	\begin{center}			
		\includegraphics[clip, width = 0.5\textwidth]{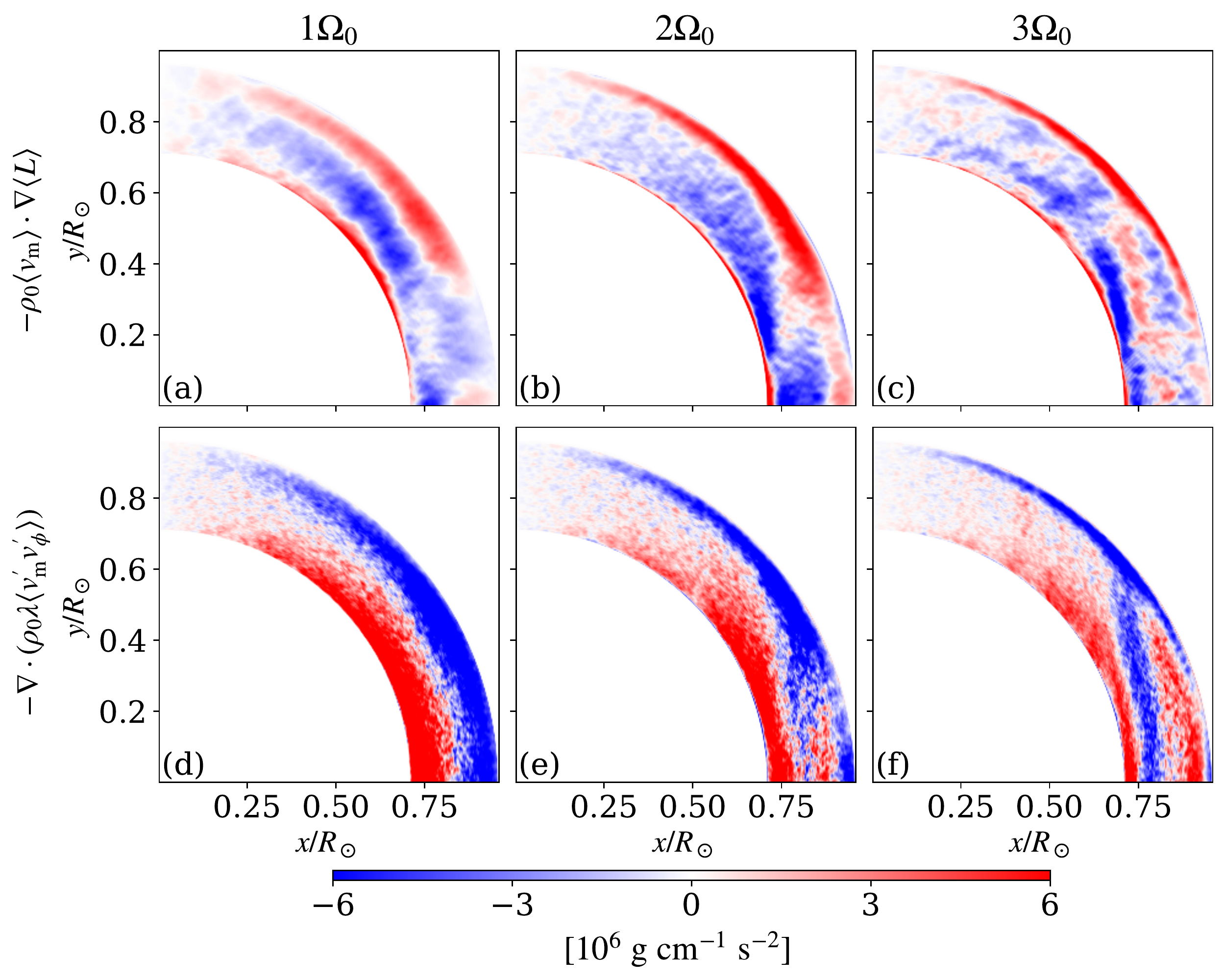}
    \caption{The AMT transport by the meridional flow: $G_\mathrm{MER}=-\rho_0\langle\bm{v}_{\mathrm{m}}\rangle\cdot\nabla\langle L \rangle$ (upper panels) and by the Reynolds stress: $G_\mathrm{REY}=-\nabla\cdot(\rho_0\lambda\langle\bm{v}_{\mathrm{m}}'v_{\phi}'\rangle)$ (lower panels) for each case are shown. The result from $1\Omega_\odot$ (panels a and d), $2\Omega_\odot$ (panels b and e), and $3\Omega_\odot$ (panels c and f) cases are shown. The two terms are mostly balanced, and the residual can be explained by the magnetic contribution.}
	\label{meridional_and_Reynolds_all}
	\end{center}
\end{figure}

Gyroscopic pumping is a useful idea for understanding the AMT balance. The gyroscopic pumping is derived from the conservation law of the angular momentum in a steady state ($\partial/\partial t=0$) \citep{miesch_2011ApJ...743...79M}, written as:
\begin{align}
	0 \sim
	\underbrace{-\rho_0\langle\bm{v}_{\mathrm{m}}\rangle\cdot\nabla\langle \mathcal{L}\rangle}_{G_\mathrm{REY}}
	\underbrace{-\nabla\cdot(\rho_0\lambda\langle\bm{v}_{\mathrm{m}}'v_{\phi}'\rangle)}_{G_\mathrm{MER}}.
\end{align}
We again ignore the magnetic field just for simplicity, although it contributes to this balance. We wish to understand the role of turbulence in the AMT. $G_\mathrm{REY}$ and $G_\mathrm{MER}$ describe the AMT by the Reynolds stress (turbulence) and the MF, respectively. Fig. \ref{meridional_and_Reynolds_all} shows $G_\mathrm{REY}$ (upper panels) and $G_\mathrm{{MER}}$ (bottom panels) in the cases.
The two terms are almost balanced in all cases. The residual of the sum of the two terms can be explained by the magnetic field. In $1\Omega_\odot$ and $2\Omega_\odot$ cases (panels d and e), the Reynolds stress $G_\mathrm{REY}$ reduces (increases) the angular momentum in the upper (lower) part of the convection. This result indicates the radially inward AMT. In the $3\Omega_\odot$ case, we see a positive contribution of $G_\mathrm{REY}$ in the low latitudes outside the tangential cylinder ($\lambda>r_\mathrm{min }$). This is the direct reason for the fast equator in the DR and multi-cell MF in the $3\Omega_\odot$ case.
\par
\begin{figure}
	\begin{center}			
		\includegraphics[width = 0.5\textwidth]{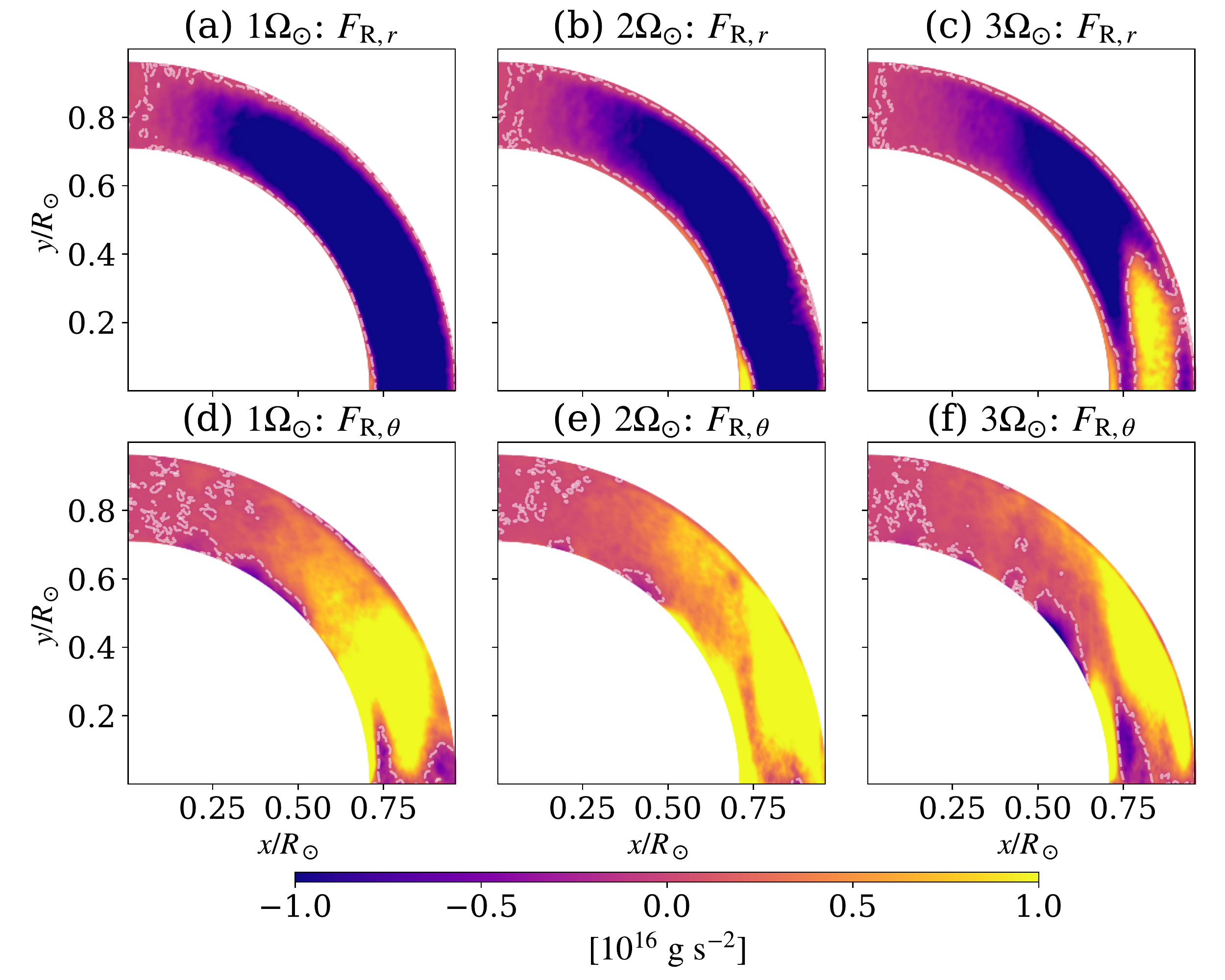}
    \caption{The AMFs by the Reynolds stress $F_{\mathrm{R},\alpha}=\rho_0\lambda\langle v_{\alpha}'v_{\phi}'\rangle$ are shown.
	The upper and lower panels show the radial and latitudinal components of the AMF, respectively. The result from $1\Omega_\odot$ (panels a and d), $2\Omega_\odot$ (panels b and e), and $3\Omega_\odot$ (panels c and f) cases are shown. The white dashed lines indicate the value of zero.
    In the $3\Omega_{\odot}$ case (panel c), we see the radially outward AMT in the low latitude, whereas the other cases show only radially inward AMT (panels a and b).}
	\label{angflux_all_all}
	\end{center}
\end{figure}

Fig. \ref{angflux_all_all} shows the AMF by the Reynolds stress $F_{\mathrm{R},\alpha}=\rho_0\lambda\langle v'_\alpha v'_\phi\rangle$.
When we reproduce the solar-like DR, i.e. the $3\Omega_\odot$ case, the radially outward AMT can be observed in the low latitude (Fig. \ref{angflux_all_all}c), which drives the multi-cell MF and the fast equator. In the other case, only the radially inward AMT is achieved in all the latitudes (Fig. \ref{angflux_all_all}a and b).
On the other hand, there is no significant difference in the latitudinal transport $F_{\mathrm{R},\theta}$ in each case (Fig. \ref{angflux_all_all}d, e, and f), i.e. all the cases show equatorward AMT. These Reynolds stress distributions are consistent with \cite{karak_2015A&A...576A..26K}; similarly in \cite{karak_2015A&A...576A..26K}, the radial Reynolds stress transports inward in almost all regions in the anti-solar type, while outward transport occurs near the equator in the solar-type. As for the latitudinal transport, both the anti-solar type and solar-type commonly transport angular momentum to the equator.
From these simulation results, we can confirm that the radial AMF $F_{\mathrm{R}, r}$ is essential for determining whether there is anti-solar- or solar-type DR.

\subsection{Scale-dependent angular momentum flux}
\label{sec:result_2}
In the previous subsection, we investigate the overall property of the AMT by the Reynolds stress (Figs. \ref{meridional_and_Reynolds_all} and \ref{angflux_all_all}). We find that the radially outward AMF is achieved in the $3\Omega_\odot$ case, where solar-like DR is reproduced. The next key question is how the AMF flux depends on the scale. Is the AMF positive in the low latitude in the $3\Omega_\odot$ on all the scales? As explained in section \ref{sec:analysis}, we decompose the AMF to four scales (see also Table \ref{ta:l}). We note that the index $i$ of $f_{\mathrm{R}, \alpha}^i$ does not represent the Fourier wavenumber $m$, but is a value defined for convenience in future discussions (see section \ref{sec:analysis} for detailed $f_{\mathrm{R},\alpha}^i$ definition).\par
\begin{figure}
	\begin{center}			
		\includegraphics[width = 0.5\textwidth]{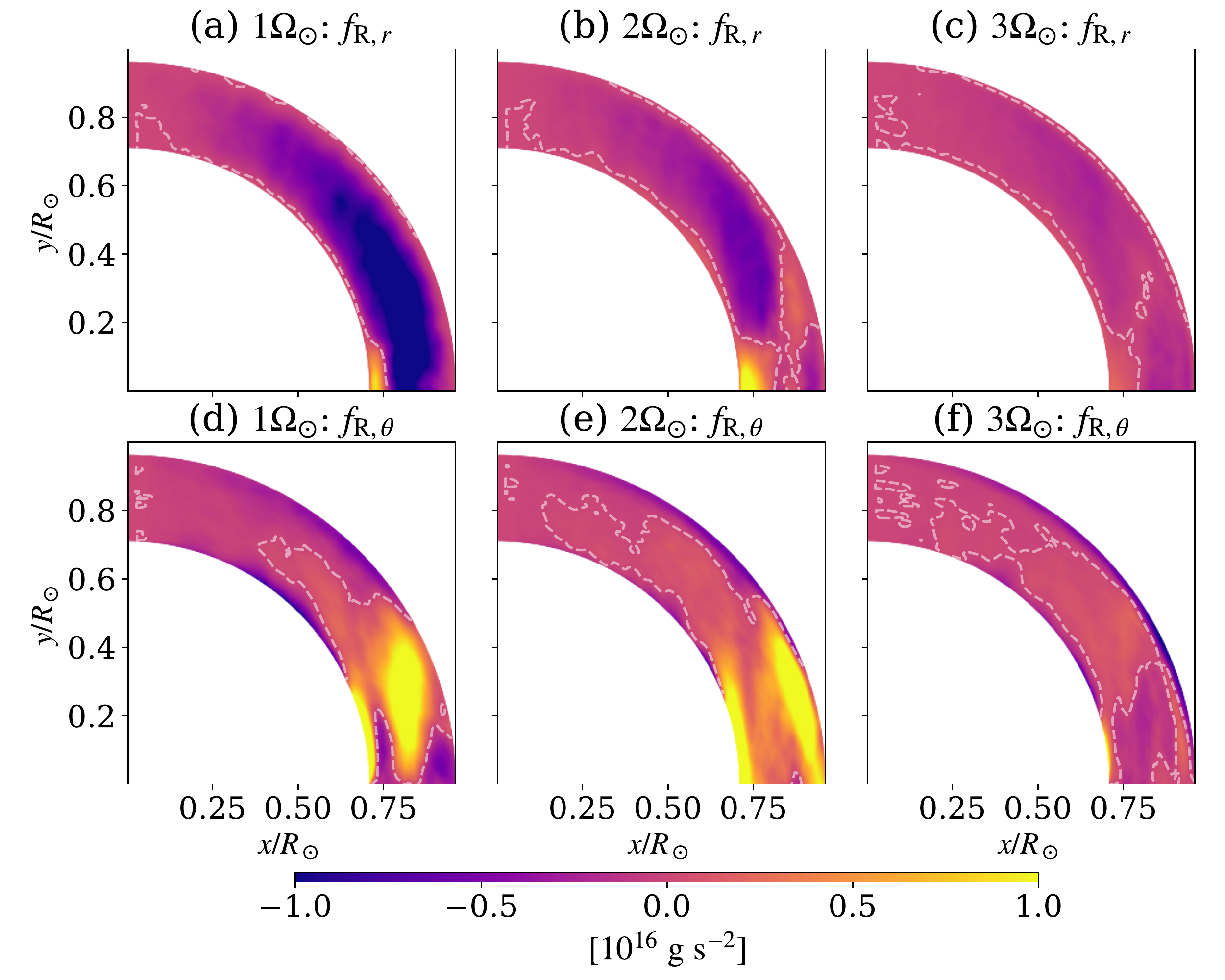}
    \caption{Scale-dependent AMFs $f^1_{\mathrm{R},\alpha}$, which correspond to the scale of $L_m\ge240~\mathrm{Mm}$ are shown. The format of the panels is identical to Fig. \ref{angflux_all_all}.
    A Gaussian filter with five grid points width is applied in all directions to reduce the realization noise.
    $f_{\mathrm{R}, r}^1$ is mostly negative in all the cases. In particular, the $1\Omega_{\odot}$ case shows strong inward transport. We cannot observe any significant AMT in the $3\Omega_\odot$ case.}
	\label{angflux_four_one}
	\end{center}
\end{figure}

Fig. \ref{angflux_four_one} shows the $f^1_{\mathrm{R},\alpha}$, which corresponds to the spatial scale of $L_m\ge240~\mathrm{Mm}$. In Fig. \ref{angflux_four_one} and the following figures, we use the same colour scale as Fig. \ref{angflux_all_all} for comparison purposes.
At this spatial scale, the radial AMT is mostly negative in all cases.
In particular, there is strong inward transport in the $1\Omega_{\odot}$ case (see Fig. \ref{angflux_four_one}a). We cannot see significant AMT in the $3\Omega_\odot$ case (Fig. \ref{angflux_four_one}c and f). This absence may be related to the suppression of the convection velocity on this scale (Fig. \ref{spectrum}).
As for the latitudinal transport $f^1_{\mathrm{R},\theta}$, the $3\Omega_{\odot}$ case has smaller values than the other two cases, but the sign is almost always positive in all the cases.\par
\begin{figure}
	\begin{center}			
		\includegraphics[width = 0.5\textwidth]{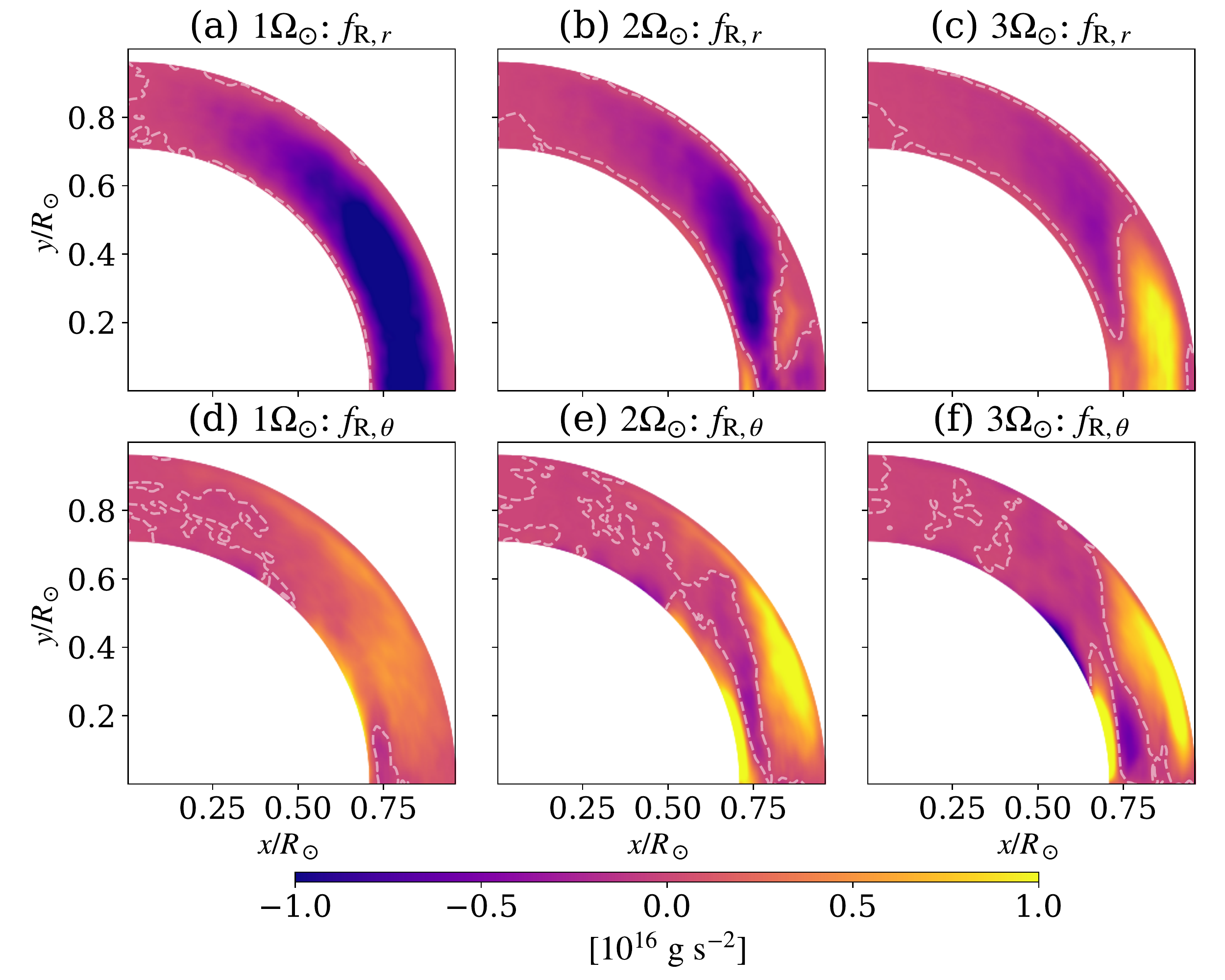}
    \caption{The same figure as Fig. \ref{angflux_four_one}, but for $f^2_{\mathrm{R},\alpha}$, which corresponds to the scale of $120~\mathrm{Mm}\le L_m<240~\mathrm{Mm}$ is shown. We can find the radially outward AMT outside the tangential cylinder in the $2\Omega_\odot$ and $3\Omega_\odot$ cases.}
		\label{angflux_four_two}
	\end{center}
\end{figure}

Fig. \ref{angflux_four_two} shows $f^2_{\mathrm{R},\alpha}$, whose spatial scale corresponds to $120~\mathrm{Mm}\le L_m<240~\mathrm{Mm}$.
At this spatial scale, radially outward transport exists outside the tangential in the $2\Omega_{\odot}$ and $3\Omega_{\odot}$ cases (Fig. \ref{angflux_four_two}b and c). It is worth noting that although we observe radially inward (negative) scale-integrated AMF $F_{\mathrm{R},r}$ in the $2\Omega_\odot$ case (Fig. \ref{angflux_all_all}b), we surely see radially outward (positive) AMF on a certain scale. In the $1\Omega_{\odot}$ case, the radially inward transport is dominant (Fig. \ref{angflux_four_two}a).
As for the latitudinal AMF $f^2_{\mathrm{R},\theta}$, the direction is always equatorward, whereas the amplitude of the AMF increases from $1\Omega_\odot$ to $3\Omega_\odot$ cases. Fig. \ref{spectrum} shows that the convective energy in this scale is smaller in the $3\Omega _\odot$ case. Thus, the increase of the amplitude in $f^2_{\mathrm{R},\theta}$ can be explained with much higher anisotropy in the turbulence (rotational influence) in the $3\Omega_\odot$ case.\par
\begin{figure}
	\begin{center}			
		\includegraphics[width = 0.5\textwidth]{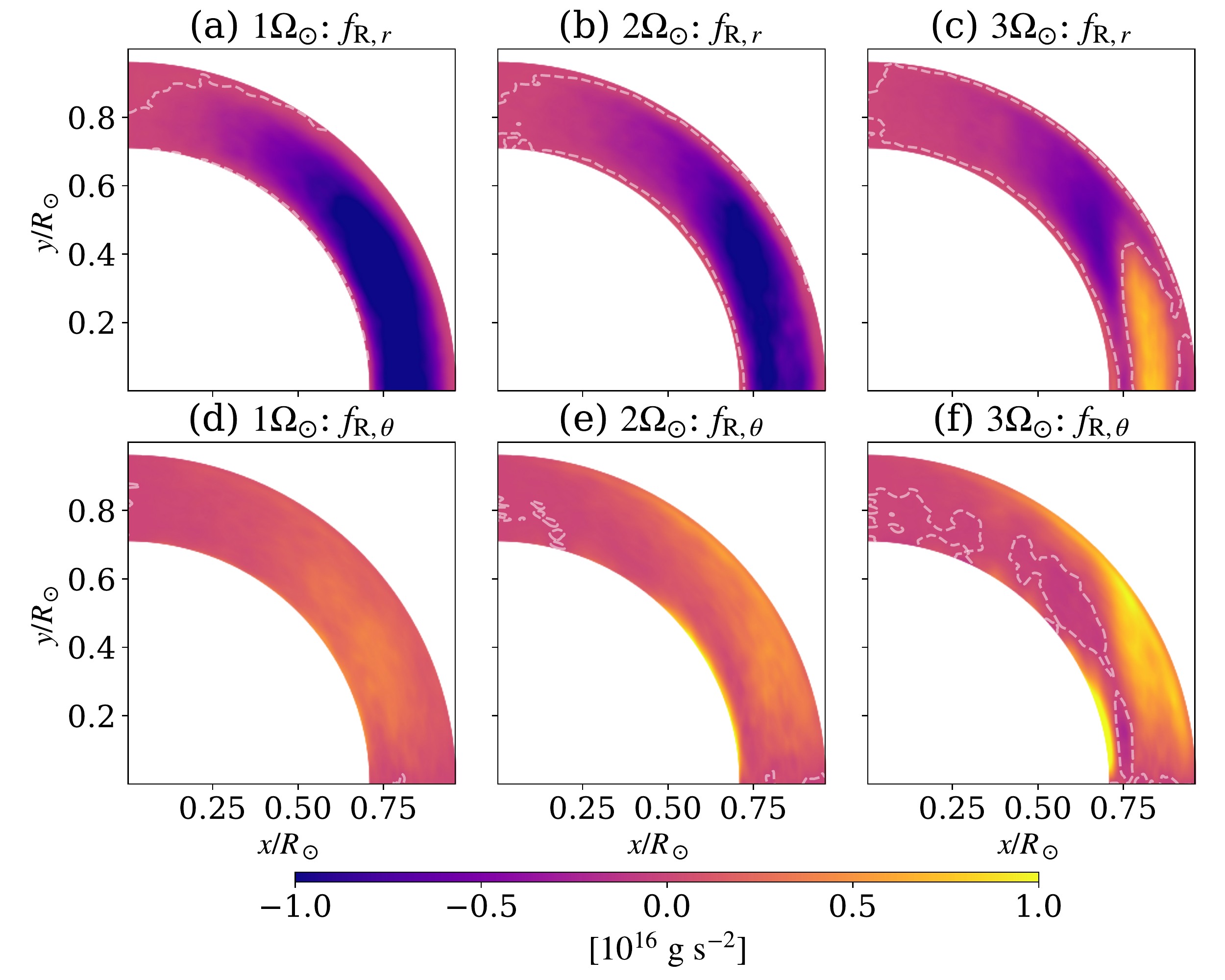}
    \caption{The same figure as Fig. \ref{angflux_four_one}, but for $f^3_{\mathrm{R},\alpha}$, which corresponds to the scale of $60~\mathrm{Mm}\le L_m<120~\mathrm{Mm}$.
	Only the $3\Omega_{\odot}$ case shows radially outward AMF outside the tangential cylinder (panel c).}
	\label{angflux_four_thr}
	\end{center}
\end{figure}

Fig. \ref{angflux_four_thr} shows the results of $f_{\mathrm{R}, \alpha}^3$. The corresponding spatial scale is $60~\mathrm{Mm}\le L_m<120~\mathrm{Mm}$.
At this spatial scale, only the $3\Omega_{\odot}$ case shows the radially outward transport $f^3_{\mathrm{R},r}$ outside the tangential cylinder (Fig. \ref{angflux_four_thr}c). The amplitude of the radially outward AMF in the $3\Omega_\odot$ decreases from $f^2_{\mathrm{R},r}$ (Fig. \ref{angflux_four_two}c, $120~\mathrm{Mm}\le L_m<240~\mathrm{Mm}$) and the negative region increases. This is consistent with our original expectation that the smaller-scale turbulence is less influenced by the rotation and tends to show a negative radial AMF. Strong radially inward AMF is dominant in the $1\Omega_{\odot}$ and $2\Omega_{\odot}$ cases (Fig. \ref{angflux_four_thr}a and b). In these two cases, the effect of rotation is smaller, i.e. the inertia term seems dominant, and the radially inward AMT is reproduced.
As for the latitudinal transport $f^3_{\mathrm{R},\theta}$, the amplitude in the $3\Omega_{\odot}$ case is larger than in the other two cases, but positive regions dominate in all cases (Fig. \ref{angflux_four_thr}d, e, and f). This tendency does not change from $f^2_{\mathrm{R},\theta}$.
\par
\begin{figure}
	\begin{center}			
		\includegraphics[width = 0.5\textwidth]{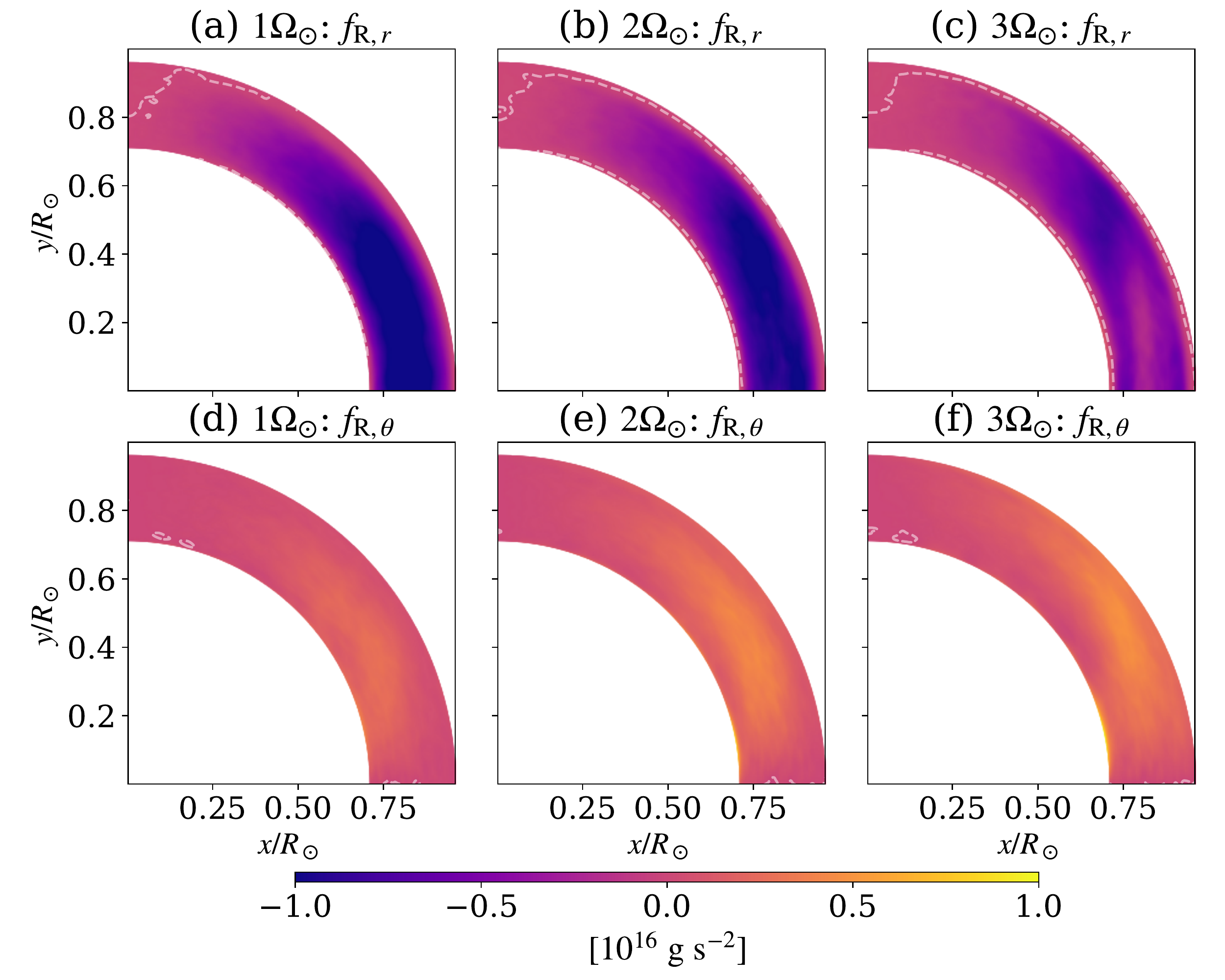}
    \caption{The same figure as Fig. \ref{angflux_four_one}, but for $f^4_{\mathrm{R},\alpha}$, which corresponds to the scale of $L_m<60~\mathrm{Mm}$. The radially inward AMF dominates in all the cases.}
	\label{angflux_four_fou}
	\end{center}
\end{figure}

The results of $f_{\mathrm{R}, \alpha}^{4}$ are shown in Fig. \ref{angflux_four_fou}. The corresponding spatial scale is $L_m<60~\mathrm{Mm}$, which is the smallest scale.
On this spatial scale, the outward radial transport of the $3\Omega_{\odot}$ case is not observed, and inward radial transport is dominant in all cases (see Fig. \ref{angflux_four_fou}a, b, and c). This result, that the radially inward transport is dominant in the smallest scale in all cases, is also consistent with our original expectation. The latitudinal AMF is also positive in all cases. The amplitude of the negative $f^3_{\mathrm{R},r}$ is smaller in the $3\Omega_\odot$ case than in the others.
\section{Summary and Conclusion}
\label{sec:conclusion}
We investigate the dependence of the AMT on the spatial scale. The AMT has a vital role in constructing large-scale effects such as DR and MF. We carry out three simulations with the different rotation rates of $\Omega_0=1\Omega_\odot, 2\Omega_\odot$, and $3\Omega_\odot$ with solar stratification. The $3\Omega_\odot$ case has a solar-like DR (fast equator) and the others have anti-solar DR (fast pole). We propose a method based on Parseval's theorem (Fourier transform) to decompose the AMF by the Reynolds stress. In our method, similar spatial scales are summed because the wavenumber $m$ in the longitudinal direction does not have a one-to-one correspondence with the spatial scale. Although magnetic fields are included in our simulations, they are not included in our analysis because Fig. \ref{meridional_and_Reynolds_all} shows that turbulence is mostly balanced with the mean flow. \par
Our result is summarized below. For the $1\Omega_\odot$ case: The rotational influence is expected to be weak. The radially inward AMT is dominant on all scales. For the $2\Omega_\odot$ case: Although the scale-integrated radial AMF is negative (radially inward, Fig. \ref{angflux_all_all}b), we observe a positive radial AMF on the scale $120~\mathrm{Mm}\le L_m<240~\mathrm{Mm}$. This indicates that while the rotational influence is weak in the $2\Omega_\odot$ as a whole, we can extract a highly rotationally constrained scale. For the $3\Omega_\odot$case: The scale-integrated radial AMT is positive (radially outward, Fig. \ref{angflux_all_all}c). The same sign for the radial AMF is seen in $60~\mathrm{Mm}\le L_m<240~\mathrm{Mm}$. Even in the $3\Omega_\odot$ case, the small scale ($L_m<60~\mathrm{Mm}$) shows a negative radial AMF, indicating the weak influence of the rotation. The strongest AMF can be seen on the scale of $120~\mathrm{Mm}\le L_m<240~\mathrm{Mm}$, where the convective energy has a peak (Fig. \ref{spectrum}). Our result suggests that the rotation influence almost disappears on the small scale. This inward transport ($\langle v_r'v_{\phi}'\rangle<0$) is thought to be formed by the strong downflow ($v_r < 0$) that dominates at small scales.
\par
We also find that the latitudinal AMF $f^i_{\mathrm{R},\theta}$ is always positive regardless of the scale. This means that the latitudinal Reynolds stress always tries to accelerate the equator but the strong radially inward AMF and the resulting MF determine the final topology of the DR, i.e. the solar-like or anti-solar DR.
\par
The most important finding in this study is that the small-scale turbulence tends to transport the angular momentum radially inward and causes the one-cell MF and resulting anti-solar DR. This is expected from the previous high-resolution simulation \citep[e.g.][]{hotta_2015ApJ...798...51H}. It is known that the high-resolution simulations tend to fall into the anti-solar DR \citep[see also][]{hotta_2022ApJ...933..199H}. This is consistent with our new finding that the small scale causes the anti-solar DR.
\par
In this study, we only focus on the AMT by the Reynolds stress (turbulence). Recently \cite{hotta_2022ApJ...933..199H} find that the magnetic field (Maxwell stress) has a vital role in reproducing the solar-like DR with the solar parameters. We can use the same method as this study to understand the scale-dependent Maxwell stress in the future.

\section*{Acknowledgements}
H.Hotta. is supported
by JSPS KAKENHI grants No. JP20K14510,
JP21H04492, JP21H01124, JP21H04497, and
MEXT as a Program for Promoting Researches
on the Supercomputer Fugaku (Toward a unified view of the universe: from large-scale
structures to planets, grant no. 20351188).
The results were obtained using the Supercomputer Fugaku provided by the RIKEN Center
for Computational Science.

\section*{Data Availability}
The simulation data underlying this article will be shared on reasonable request to the corresponding author.
 



\bibliographystyle{mnras}
\bibliography{diffrot_1_mnras_re2_all_final} 




\appendix

\bsp	
\label{lastpage}
\end{document}